\documentclass[%
 aip,
 amsmath,amssymb,
 reprint,%
]{revtex4-1}

\usepackage{graphicx}
\usepackage{dcolumn}
\usepackage{bm}

\usepackage[utf8]{inputenc}
\usepackage[T1]{fontenc}
\usepackage{mathptmx}
\usepackage{etoolbox}

\usepackage{amssymb}
\usepackage{float}

\usepackage{hyperref}
\usepackage{xcolor}
\usepackage{nameref}

\usepackage{graphicx}
\usepackage{subcaption}
\usepackage{epstopdf}

\makeatletter
\def\@email#1#2{%
 \endgroup
 \patchcmd{\titleblock@produce}
  {\frontmatter@RRAPformat}
  {\frontmatter@RRAPformat{\produce@RRAP{*#1\href{mailto:#2}{#2}}}\frontmatter@RRAPformat}
  {}{}
}%
\makeatother
\begin{document}

\preprint{AIP/123-QED}

\title[]{Wake dynamics of a square cylinder while moving upward in quiescent water}
\author{Intesaaf Ashraf}
\email{intesaaf.ashraf@ucl.ac.uk}
\altaffiliation[]{Physics Department, XYZ University, Université de Liège, Liege - 4000, Belgium}

\author{Neetu Tiwari}
\affiliation{Mechanical \& Aerospace Engineering, Indian Institute of Technology Hyderabad, Sangareddy - 502285, Telangana, India}

\author{Stephane Dorbolo}%
\affiliation{ 
Physics Department, XYZ University, Université de Liège, Liege - 4000, Belgium
}%
\date{\today}

\begin{abstract}
We experimentally investigate the wake dynamics of a square cylinder rising through quiescent water over a range of Froude numbers $(\mathrm{Fr})$. Time-resolved Particle Image Velocimetry provides velocity and vorticity fields that enable pressure reconstruction and vortex characterization. Diagnostics based on swirl strength $(\lambda_{ci})$, the Okubo--Weiss parameter $(W)$, and a shear--vortex interaction measure $(\Lambda)$ reveal that the wake is governed by a persistent pair of counter-rotating vortices rather than by periodic shedding. Circulation exhibits a two-regime dependence on $\mathrm{Fr}$, with a sharp increase below $\mathrm{Fr}\approx 1$ and saturation above this threshold, mirroring entrainment force scaling reported previously. While vortex area remains nearly constant, swirl strength and negative-$W$ regions expand with $\mathrm{Fr}$, indicating that entrainment enhancement arises from intensified rotation rather than an enlarged vortex footprint. These findings provide new physical insight into vortex–free-surface interactions and enrich the understanding of entrainment mechanisms in unsteady wakes, with implications for multiphase flows and the hydrodynamic design of naval and offshore structures.
\end{abstract}

\maketitle

\maketitle

\section{Introduction}\label{sec1}

The interaction of rigid bodies with free water surfaces during entry or exit processes is crucial across diverse engineering disciplines, such as naval architecture, ocean and coastal engineering, aviation, and dip-coating technologies. Due to the inherent complexities involving fluid-structure interactions, interfacial dynamics, and fluid mechanics, these phenomena remain challenging to interpret and accurately model.

Extensive research has explored the physics of bodies entering~\citep{challa2014,challa2010,mohtat2015,yang2012} and exiting fluids~\citep{Havelock1936,truscott2016,wu2017experimental,haohao2019numerical}. Recent studies have increasingly focused on water exit scenarios to deepen the understanding of associated flow dynamics, validate computational models, and guide the development of analytical frameworks~\citep{ashraf2024, ashraf2024effect, ashraf2024exit, takamure2025motion, huang2024numerical, zhou2024numerical}. 

Pioneering analytical work by \citet{Havelock1936} addressed the vertical motion of cylinders in uniform flow. \citet{greenhow1983nonlinear} subsequently performed experiments on a neutrally buoyant cylinder rising from a stationary position, observing characteristic bump-like free-surface deformations culminating in chaotic "waterfall breaking." Numerically, \citet{telste1987} used potential flow theory to categorize the water exit of cylinders into three regimes: low-speed "wall-like" flow, intermediate-speed wave generation, and high-speed infinite-domain-like behavior. More recent experiments by \citet{ashraf2024} revealed symmetry-breaking wake instabilities in water in contrast to symmetric wakes in viscous silicone oil. They reported a two-stage drainage behavior of the entrained fluid, initially exponential, transitioning to a power-law decrease (\(t^{-1/2}\)) at later times, alongside a clear correlation of drag and entrainment forces with the Froude number.

Other researchers have also contributed significantly to the understanding of water exit processes. For instance, \citet{greenhow1997water} performed two-dimensional simulations to investigate submerged cylinders in forced upward motion. \citet{liju2001} analyzed the surface surge phenomenon for cylinders of varying geometries and Reynolds numbers, while \citet{moshari2014} employed Volume of Fluid (VOF) simulations to capture wave formation and air entrainment during oblique water exits. Additional numerical studies utilizing advanced VOF techniques were conducted by \citet{kleefsman2004improved} and \citet{nair2018water}. Experimental work by \citet{chu2010} identified cavity formation at cylinder extremities, resulting in water slapping upon cavity collapse during both accelerating and decelerating motions.

In the context of spherical bodies, numerical analyses by \citet{haohao2019numerical}, employing Lattice Boltzmann methods (LBM), demonstrated a strong dependency of water-surface elevation on the Froude number, notably identifying intensified waterfall breaking at higher velocities. Furthermore, \citet{ni2015} observed delayed free-surface detachment for blunter spheroids. Experimentally, \citet{truscott2016} underscored the significance of the Reynolds number in determining vortex shedding and trajectories of buoyant spheres rising toward the surface, with distinct linear or oscillatory pathways based on initial depth. Complementary findings by \citet{wu2017experimental} showed velocity-dependent surface elevations for fully submerged spheres, contrasting sharply with distinct columnar water detachments for partially submerged cases. Recently, comparative experimental analyses demonstrated that spheres with surface dimples significantly reduce drag and entrainment force coefficients at higher Froude numbers, enhancing water-exit efficiency, though leaving cross-over forces largely unaffected~\citep{ashraf2024effect}.

\textcolor{black}{The hydrodynamic behaviour of a rising square cylinder shares several universal  features with previously studied water-exit flows of smooth bodies such as  spheres and circular cylinders, while also exhibiting geometry-specific 
differences. In all cases, the exit process follows two principal stages: an inertia-dominated bulging of the free surface, followed by a drainage-limited phase in which the entrained liquid volume and wake vorticity gradually decay 
once the body crosses the interface~\cite{coutanceau1977experimental, bouard1980early, ashraf2024exit, ashraf2024, moshari2014}. 
Our recent study on the inertial exit dynamics of a square cylinder~\cite{ashraf2024exit}  demonstrated that, despite geometry-specific shear-layer effects, the global entrainment and circulation trends collapse with those of circular cylinders when normalized by projected area and Froude number. The present flow-field observations extend that analysis by directly resolving the unsteady vortex 
dynamics governing these entrainment regimes. These results are consistent with previous numerical predictions of buoyancy-driven exits, which report less than a few percent deviation between two- and three-dimensional simulations of interface evolution~\cite{moshari2014}, and with experimental studies showing similar entrained-volume and force scaling across different body  shapes~\cite{ashraf2024}. Overall, despite the enhanced local shear associated with the square geometry, the large-scale momentum redistribution and two-stage exit dynamics remain broadly consistent with those observed for smoother geometries, situating the present work within the broader framework of non-axisymmetric water-exit flows.}

\textcolor{black}{In this study, the exit dynamics of a square cylinder withdrawn vertically from a quiescent water surface at constant speed are investigated. The evolution of the near-wake vortex structures and their interaction with the deforming free surface are quantified using time-resolved particle image velocimetry (PIV). Two dimensionless parameters are used to describe the flow: the Froude number is defined as, \begin{equation}
Fr = \frac{U^2}{ga}
\end{equation}, where $U$ is the withdrawal velocity and $a = 0.04m$ is the cylinder side length.  The Reynolds number is defined as
\begin{equation}
Re = \frac{U a}{\nu},
\end{equation}
where $U$ is the withdrawal velocity, $a$ is the characteristic side length of the square cylinder, and $\nu$ is the kinematic viscosity of water ($\nu \approx 1.0\times10^{-6}\,\mathrm{m^2/s}$ at $20^{\circ}\mathrm{C}$).
For the range of velocities investigated ($U = 0.2$--$1.0\,\mathrm{m/s}$) and a cylinder side length of $a = 0.04\,\mathrm{m}$, the corresponding Reynolds numbers span
\begin{equation}
Re = \frac{U a}{\nu} \approx (8\times10^{3}\text{--}4\times10^{4}).
\end{equation}
The analysis focuses on the vortex formation, circulation dynamics, and entrained-liquid evolution across a range of $Fr$ and $Re$, thereby elucidating the fluid-dynamic mechanisms underlying the two observed entrainment regimes.}

\section{Methodology}
\subsection{\label{setup}Experimental Set--up}

The experimental setup comprised a lifting mechanism, a glass water tank (dimensions: 78.5 cm × 27.5 cm × 72.5 cm), PIV, and the test object. The lifting mechanism utilized a rack-and-pinion system, with the rack’s moving part connected to a carbon fiber frame. This frame securely held the test cylinder in a horizontal orientation, parallel to the water tank’s side and front walls. The test object was a right square cylinder, 3D-printed with square bases of side \( a = 40 \, \text{mm} \) and \textcolor{black}{length, $L = 220$mm.} \textcolor{black}{Although the present square cylinder has a finite aspect ratio ($L/a \approx 5.5$), the classical criterion of $L/D \gtrsim 10$ for quasi-two-dimensional wakes~\cite{coutanceau1977experimental,bouard1980early} applies primarily to uniform cross-flow past fixed cylinders, where strong three-dimensional vortex dislocations occur along the span. 
In contrast, the current configuration involves a vertically accelerating cylinder rising from rest in quiescent water, for which the wake is buoyancy-driven and symmetric about the mid-plane. Under such conditions, end-effects are substantially weaker, and both experimental and numerical studies of similar buoyancy-driven exits~\cite{ashraf2024} have shown less than 2\% deviation 
between two- and three-dimensional predictions at $L/a \approx 5.5$. Consistently, the present PIV measurements exhibit no measurable spanwise asymmetry, indicating that the extracted two-dimensional fields faithfully represent the central-span dynamics.}

\textcolor{black}{The cylinder was initially submerged to an immersion depth \(H = 326~\mathrm{mm}\) (\(H/a = 8.15\)). During upward motion, the instantaneous immersion depth is expressed in non-dimensional form as \(h = H_i/a\), where \(H_i\) is the instantaneous distance from the free surface to the cylinder centre. The cylinder was aligned horizontally to ensure that its square bases remained perpendicular to the water surface at all times. The experimental setup has been explained in detail in the authors' previous work \ citep {ashraf2024exit}.}

\begin{figure*}
	\centering
	\includegraphics[scale=0.45]{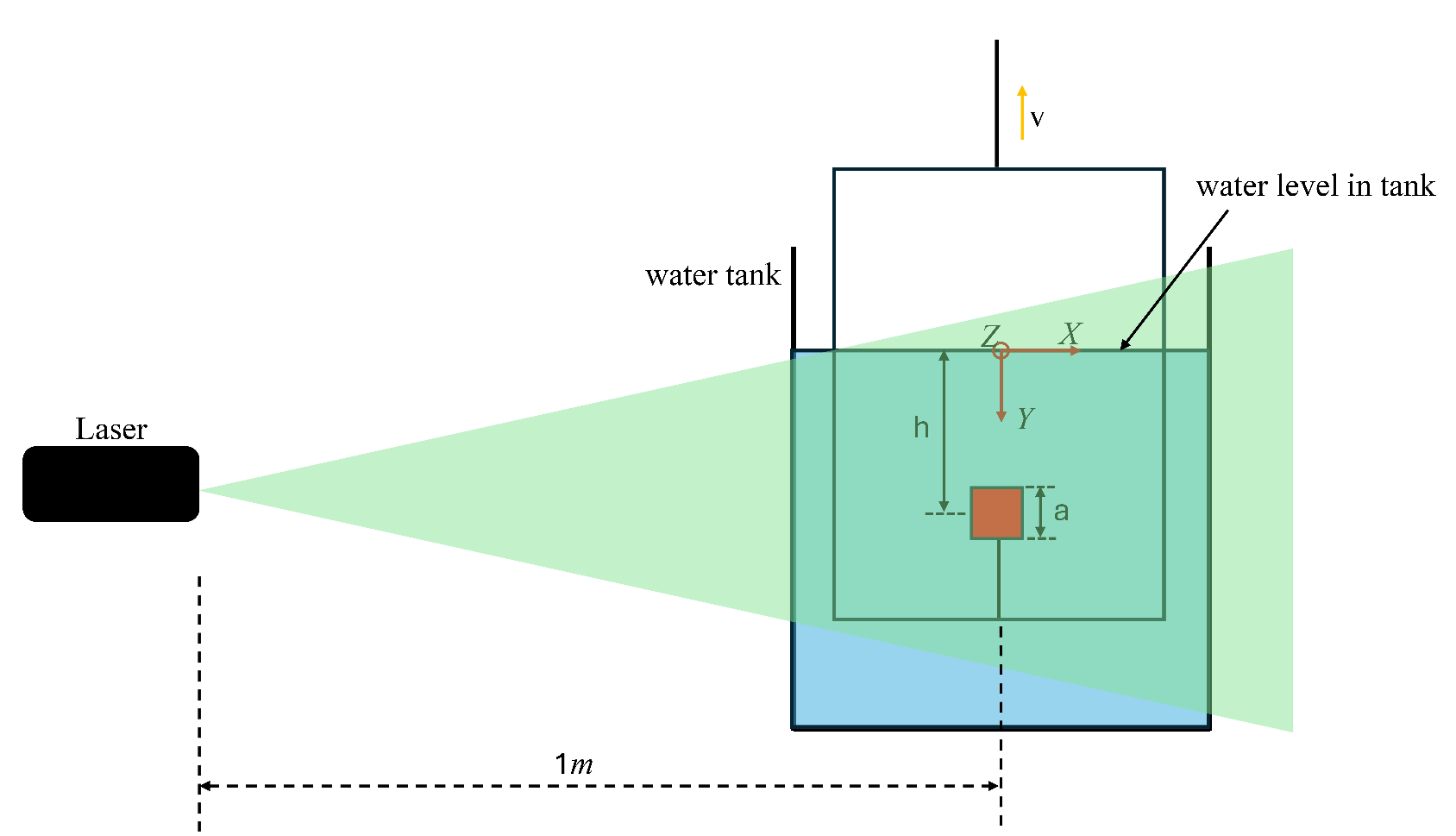}
	\caption{ Schematic of the experimental configuration with the laser positioned 1 meter away from the cylinder’s center. Image acquisition was performed at a frequency of 2000 Hz (figure is not to scale). Here, \(h\) is the instantaneous non-dimensional immersion depth (\(h = H_i/a\)) during the rise.  Coordinates \((X,Y)\) are normalized by the cylinder side length \(a\).
}

	\label{PIV_setup}
\end{figure*}

The 2D planar Particle Image Velocimetry (PIV) technique was employed for the flow measurements. A laser with a \( 532 \, \text{nm} \) wavelength and \( 4 \, \text{W} \) peak power illuminated tracer particles of approximately \( 20 \, \mu \text{m} \) in diameter. Images were recorded at \( 2000 \, \text{fps} \) (frames per second) with an interrogation window size of \( 32 \times 32 \, \text{pixels} \), a magnification factor of \( 6.075 \, \text{pixels/mm} \), and a field of view measuring \( 0.128 \, \text{m} \times 0.205 \, \text{m} \). The laser plane was aligned with the \( xy \)-plane at a distance of \( 1 \, \text{m} \) from the cylinder’s \textcolor{black}{mid section.} \textcolor{black}{The accuracy of the time-resolved PIV measurements was assessed in terms of both particle-image displacement and velocity-field uncertainty. The mean particle displacement within each interrogation window was approximately $0.9$~pixels per frame (equivalent to $\sim0.15~\mathrm{mm/frame}$ at a spatial calibration of 40~mm per 245~pixels), with a median of $0.6$~pixels and a range of $0.01$--$5.3$~pixels. The small mean displacement reflects the use of a high acquisition rate ($2000$~Hz), which enables resolution of the rapid transient motion around the rising cylinder while minimizing peak-locking errors. Regions of negligible displacement correspond to quiescent flow zones outside the wake, whereas larger displacements occur near the shear layer and within the entrained vortex cores. No additional high-frequency filtering, such as proper orthogonal decomposition (POD), was applied to the velocity fields. Instead, the data were smoothed within \textit{PIVlab} using its built-in local spatial and temporal filters, which effectively suppress spurious vectors and sub-pixel noise without altering the physical frequency content of the flow. This procedure provides a mild low-pass effect consistent with the measurement resolution and sampling frequency. The velocity uncertainty, $E_U$, was quantified using two independent approaches: (i) the signal-to-noise-ratio (SNR) correlation method of \citet{Wieneke2015}, and (ii) the particle-disparity technique of \citet{sciacchitano2013piv}. Both methods yielded consistent results, with typical absolute uncertainties of $E_U = 0.05$--$0.08~\mathrm{m/s}$, corresponding to approximately 3--5\% of the local instantaneous velocity magnitude. To assess the impact on spatial derivatives, this uncertainty was propagated through the grid spacing ($\Delta x \simeq 5.2~\mathrm{mm}$) to estimate a characteristic
uncertainty in the velocity-gradient tensor:
\begin{equation}
E_{\partial u / \partial x} \sim \frac{E_U}{\Delta x} \approx 10~\mathrm{s^{-1}},
\end{equation} which is at least an order of magnitude smaller than the characteristic strain-rate and vorticity values ($>100~\mathrm{s^{-1}}$) observed in the wake. Consequently, the uncertainty has a negligible influence on the interpretation of the velocity-gradient tensor components and derived quantities such as vorticity and circulation. The temporal evolution of the velocity and vorticity fields is presented in Figures~2--3 for three representative Froude numbers, and further detailed frame-by-frame in Appendix~A (Figures~ \ref{V200}--\ref{V800_0}). The observed temporal variations are well above the estimated uncertainty bounds, confirming that the unsteady flow dynamics are fully resolved within the measurement accuracy of the present PIV setup.}

The experimental setup for PIV measurements is depicted in Fig. \ref{PIV_setup}. The acquired PIV images were post-processed using the open-source software PIVLab \citep{thielicke2014pivlab}. The coordinate system (\( XYZ \)) used in the experiment is shown in Fig. \ref{PIV_setup}. The \( X \)-axis runs along the length of the tank, the \( Z \)-axis along its width, and the \( Y \)-axis represents the vertical direction, aligned with the pulling motion. The origin is set at the intersection between the pulling axis and the free water surface, defined as the horizontal plane (\( XZ \)). All coordinates were normalized by dividing them by \( a \), the side length of the cylinder’s square bases.

The experiments were performed at constant terminal velocities of  0.2 \text{m/s}, 0.25 \text{m/s}, 0.3 \text{m/s}, 0.4 \text{m/s}, 0.6 \text{m/s}, 0.8 \text{m/s}  and  1 \text{m/s}, with a fixed acceleration of \( 4 \, \text{m/s}^2 \). To reach the fixed terminal velocity, the cylinder has to travel a predetermined distance from rest, such as \( 120 \, \text{mm} \) to achieve \( 1 \, \text{m/s} \). The same conditions applied during the deceleration phase. The motion of the cylinder was verified to maintain constant speed during the PIV measurements.


\subsection{Pressure Field Estimation from PIV}
Two methods have generally been used for pressure estimation from PIV-measured velocity data. One of them is solving the Navier--Stokes equations via path integration \citep{tiwari2019pressure}; the other is solving the pressure Poisson equation \citep{tiwari2021ultrasonic, tiwari2021piv}. The pressure can be estimated from the Navier--Stokes equations as follows:

\begin{equation}
\nabla p= -\rho \frac{D\mathbf{u}}{Dt} + \mu \nabla ^ 2\mathbf{u}
\label{equ1}
\end{equation}

where \(\mathbf{u}\) is the fluid flow velocity, \(p\) is the pressure, \(\rho\) is the density, and \(\mu\) is the local dynamic viscosity. The PIV-measured velocity data provide the information necessary to estimate the pressure gradient. In the Eulerian approach~\cite{VanDerKindere2019}, the material (or total) acceleration can be expressed as

\begin{equation}
\frac{D \mathbf{u}}{Dt}
= \frac{\partial \mathbf{u}}{\partial t}
+ (\mathbf{u} \cdot \nabla) \mathbf{u}.
\label{equ2}
\end{equation}

Time separation between two consecutive snapshots has a significant effect on the accuracy of the estimated pressure field. For example, if we discretize the local acceleration term using a second-order finite difference scheme, we can define the associated truncation error as below:

\begin{equation}
\frac{\partial \mathbf{u}}{\partial t} =  \frac{\mathbf{u}(x,t+\Delta t) -  \mathbf{u}(x,t - \Delta t)}{2 \Delta \mathbf{t}} +\frac{\Delta t^2}{6} \frac{\partial ^3 \mathbf{u}}{\partial t^3}
\label{equ3}
\end{equation}

where higher-order terms contributing to truncation error, $\frac{(\Delta t^2)}{6} \frac{\partial ^3 \mathbf{u}}{\partial t^3}$, are usually neglected. The velocity measurement from PIV usually has uncertainty, so we measure velocity in the range  $ \mathbf{u \pm \epsilon _\mathbf{u}}$, and the details of these uncertainty sources can be found in the literature\cite{Sciacchitano2019}. Generally, uncertainties propagate when derivative terms of velocity are evaluated. In the local acceleration term, precision errors arise due to the velocity measurement uncertainty ($\epsilon _u$))~\cite{VanOudheusden2013}:

\begin{equation}
\epsilon ^2 \frac{\partial u}{\partial t} \approx  \frac {\sqrt{{\epsilon _\mathbf{u}^2} + {\epsilon _\mathbf{u}^2}}}{4 \Delta \mathbf{t}^2}  \approx \frac{\epsilon _\mathbf{u}^2}{2 \Delta \mathbf{t}^2}
\label{equ4}
\end{equation}

The effect of time separation on the truncation and precision errors is discussed in detail in the literature~\cite{VanOudheusden2013}. With a high sampling rate, time separation can be reduced, and the truncation error can also be reduced. However, precision errors will increase (\cite{VanOudheusden2013}). Therefore, the pressure field estimated using the Navier-Stokes equations is not error-free, even with time-resolved data. Thus, for unsteady flows, the Poisson pressure equation is preferred; however, appropriate boundary conditions are critical. The Poisson pressure equation can be derived by taking the divergence of equation \ref{equ1},

\begin{equation}
\nabla ^2 p
= -  \rho \nabla . (\mathbf{u.\nabla)\mathbf{u}}
\label{equ5}
\end{equation}

In this study, Dirichlet boundary conditions were selected. \textcolor{black}{The Dirichlet boundary conditions were imposed by prescribing fixed values of velocity and pressure at the domain boundaries as
\begin{subequations}
\begin{align}
\mathbf{u} &=
\begin{cases}
\mathbf{0}, & \text{on stationary and far-field boundaries},\\
\mathbf{U}(t), & \text{on the moving cylinder surface},
\end{cases} \\[3pt]
p &= 0 \quad \text{at the free surface.}
\end{align}
\label{eq:dirichlet}
\end{subequations}
where \(\mathbf{U}_w(t)\) is the prescribed wall velocity following the imposed acceleration profile.} The pressure Poisson equation approach requires the evaluation of a second derivative of the velocity, which is not a direct measurement of PIV. The primary sources of error in the pressure Poisson equation are non-exact boundary conditions and the calculation of velocity gradients. However, by adopting specific measures during the experiments, the error can be minimized in the pressure Poisson equation (~\cite{Pan2016}). In the present study, in solving the pressure Poisson equation, the iteration was repeated until the root mean square error was reduced to the order of  $10^{-5}$.

\textcolor{black}{
The local pressure distribution was expressed using the non-dimensional pressure coefficient,
\begin{equation}
C_p = \frac{p - p_\infty}{\tfrac{1}{2}\rho U_{\!r}^2},
\label{eq:Cp}
\end{equation}
where \(p\) is the instantaneous local pressure, \(p_\infty\) is the reference hydrostatic pressure in the undisturbed fluid at the same depth, \(\rho\) is the fluid density, and \(U_{\!r}\) is the instantaneous rise velocity of the cylinder. The reference pressure is defined as \(p_\infty = p_0 + \rho g y\), where \(p_0\) is the atmospheric pressure at the free surface and $y$ is the local vertical coordinate of the fluid point measured downward from the undisturbed free surface. Subtracting \(p_\infty\) removes the static hydrostatic contribution, allowing \(C_p\) to quantify only the dynamic pressure fluctuations associated with the unsteady wake formation.
}

\subsection{Okubo--Weiss parameter}
\label{subsec:OWparameter}

To distinguish regions of shear-dominated flow from those dominated by rotation, we employ the Okubo--Weiss parameter $W$ \cite{okubo1970horizontal,weiss1991dynamics}. For a two-dimensional, incompressible velocity field $\mathbf{u}=(U,V)$, $W$ is defined as
\begin{equation}
W = s_n^2 + s_s^2 - \omega^2,
\label{equ:OW}
\end{equation}
where $\omega$ is the out-of-plane vorticity,
\begin{equation}
\omega = \frac{\partial V}{\partial x} - \frac{\partial U}{\partial y},
\label{equ:vort}
\end{equation}
and $s_n$, $s_s$ are the normal and shear strain rates. These are obtained from the velocity–gradient tensor
\begin{equation}
\nabla \mathbf{u} \;=\;
\begin{bmatrix}
\partial U/\partial x & \partial U/\partial y\\
\partial V/\partial x & \partial V/\partial y
\end{bmatrix},
\label{equ:grad}
\end{equation}
which can be decomposed into symmetric (strain) and antisymmetric (rotation) parts:
\begin{equation}
\nabla \mathbf{u} = \frac{1}{2}\big(\nabla \mathbf{u}+(\nabla \mathbf{u})^T\big)
+ \frac{1}{2}\big(\nabla \mathbf{u}-(\nabla \mathbf{u})^T\big).
\label{equ:decomp}
\end{equation}

For a 2D incompressible velocity field, the explicit components are
\begin{equation}
s_n = \frac{\partial U}{\partial x} - \frac{\partial V}{\partial y}, \qquad
s_s = \frac{\partial U}{\partial y} + \frac{\partial V}{\partial x}, \qquad
\omega = \frac{\partial V}{\partial x} - \frac{\partial U}{\partial y}.
\label{equ:snss}
\end{equation}
These enter directly into Eq.~\ref{equ:OW}.

Regions of flow where $W<0$ indicate rotation-dominated (elliptic) zones, typically associated with vortex cores. Conversely, $W>0$ signifies shear-dominated (hyperbolic) zones, often surrounding vortical structures. Thus, $W$ provides a convenient scalar measure to delineate coherent vortices from adjacent shear layers.

In practice, velocity gradients are calculated using a central-difference scheme. The resulting $W$ field is evaluated at each PIV grid point, yielding instantaneous two-dimensional maps that can be analyzed frame by frame or ensemble-averaged to characterize wake topology during cylinder exit.

Traditional diagnostics each have limitations: vorticity $\omega$ cannot distinguish between rotational and shear-dominated regions, while swirl strength $\lambda_{ci}$ identifies vortex cores but neglects surrounding shear layers. The Okubo--Weiss parameter $W$ overcomes these shortcomings by explicitly balancing strain and rotation. Regions with $W<0$ correspond to coherent vortices, whereas $W>0$ identifies shear-dominated zones. This discrimination is particularly important in our case, where the wake is characterized not by periodic shedding but by persistent vortices embedded in strong shear layers. Hence, $W$ provides a more physically consistent basis for quantifying the evolving shear--vortex balance.

\textcolor{black}{Alternative vortex-identification formulations such as the Liutex (or Rortex) method~\cite{liu2019third, liu2021liutex} have also been proposed to isolate purely rotational motion in three-dimensional flows. 
In the present quasi-two-dimensional configuration, however, the combined use of swirl strength ($\lambda_{\mathrm{ci}}$), Okubo--Weiss ($W$), and the shear--vortex interaction parameter ($\Lambda$) provides an equivalent and physically transparent characterization of the wake dynamics, and thus no additional Liutex-based computation was required.}

\subsection{Shear–vortex interaction parameter}
\label{sec:Lambda_def}

We further quantify the relative footprint of shear– and vortex–dominated regions using a shear–vortex interaction parameter $\Lambda$. The procedure is as follows:

\paragraph{(i) Swirl strength.}
Let $\nabla \mathbf{u}$ be the in–plane velocity–gradient tensor. Its eigenvalues are
\[
\lambda = \frac{\mathrm{Tr}(\nabla \mathbf{u})}{2} \;\pm\;
\sqrt{\left(\frac{\mathrm{Tr}(\nabla \mathbf{u})}{2}\right)^{2} - \det(\nabla \mathbf{u})},
\]
and the local swirling intensity is $\lambda_{ci} = \mathrm{Im}(\lambda)$.

\paragraph{(ii) Vortex–core mask.}
A vortex mask $M_v(\mathbf{x})=1$ where $\lambda_{ci}(\mathbf{x}) \ge \alpha \,\lambda_{ci,\max}$
and $0$ otherwise, with $\alpha=0.05$ (robust across $0.03\le\alpha\le0.08$).

\paragraph{(iii) Shear–dominated mask.}
Using the Okubo–Weiss field $W = s_n^2 + s_s^2 - \omega^2$ (Sec.~\ref{subsec:OWparameter}),
we define a shear mask $M_s(\mathbf{x})=1$ where $W(\mathbf{x})>0$ and $0$ otherwise.

\paragraph{(iv) Shear–vortex interaction parameter.}
Finally,
\begin{equation}
\Lambda \;=\;
\frac{\displaystyle \int_{\Omega} M_s(\mathbf{x})\,\mathrm{d}A}
{\displaystyle \int_{\Omega} M_v(\mathbf{x})\,\mathrm{d}A}.
\label{equ:Lambda}
\end{equation}

Thus, $\Lambda<1$ indicates vortex–dominated topology (vortex cores occupy more area than shear layers), while $\Lambda>1$ indicates shear–dominated topology. Because $M_v$ and $M_s$ are constructed from complementary diagnostics ($\lambda_{ci}$ and $W$), $\Lambda$ avoids the ambiguity of using $\lambda_{ci}$ alone to represent shear. We verified that $\Lambda(h,Fr)$ trends are insensitive to the threshold $\alpha$ within $0.03 \le \alpha \le 0.08$ and to $\pm 1$–grid shifts in gradient evaluation.

\subsection{Vortex dynamics characterization}
\label{subsec:vortex_char}

To analyze the vortical structures in the wake of the square cylinder, we employed a methodology based on detecting coherent vortex regions using the swirl strength ($\lambda_{ci}$) and vorticity ($\omega$) \cite{chen2018evaluation,chen2014improved,morton1969strength}. The velocity fields, obtained from PIV at discrete time steps, were processed as follows:

\textbf{Preprocessing:} The velocity components $(U,V)$ were smoothed using a Gaussian filter to reduce noise and ensure stable vortex detection.

\textbf{Velocity gradient computation:} The velocity field was numerically differentiated to obtain the velocity–gradient tensor using Eq.~\ref{equ:grad}.

\textbf{Swirl strength calculation:} The swirl strength $\lambda_{ci}$ was computed according to Sec.~\ref{sec:Lambda_def}.

\textbf{Vorticity calculation:} The vorticity field $\omega$ was calculated using Eq.~\ref{equ:vort}.

A combination of $\lambda_{ci}$ and $|\omega|$ was used to distinguish vortex cores from surrounding shear layers \cite{chakraborty2007kinematics}.

\textbf{Morphological processing:} To ensure spatial continuity of vortex regions, morphological closing was applied using a disk structuring element.

\textbf{Region labeling and filtering:} Connected vortex regions were identified using an 8–connectivity labeling algorithm. Only the largest clockwise (CW) and counterclockwise (CCW) vortices per frame were retained to eliminate spurious detections.

\textbf{Circulation and vortex area:} The circulation $\Gamma$ and vortex area $A$ of the largest vortex were computed as
\begin{equation}
    \Gamma = \sum_{\text{vortex region}} \omega \, \Delta x \, \Delta y,
\end{equation}
\begin{equation}
    A = \sum_{\text{vortex region}} \Delta x \, \Delta y,
\end{equation}
where $\Delta x$ and $\Delta y$ are the grid spacings in the PIV field.

The detected vortices were visualized by overlaying color–mapped vorticity contours, with red and blue corresponding to the largest CW and CCW vortices, respectively. This workflow ensured robust identification of dominant vortices and enabled quantitative tracking of circulation and vortex size evolution over time.


\section{Results and Discussion}
\label{sec:ResultsDiscussion}

\subsection{Overall wake evolution across Froude number}
\label{subsec:FlowEvolution}

\begin{figure*}
    \centering
    \begin{subfigure}[b]{0.72\textwidth}
        \includegraphics[width=\textwidth]{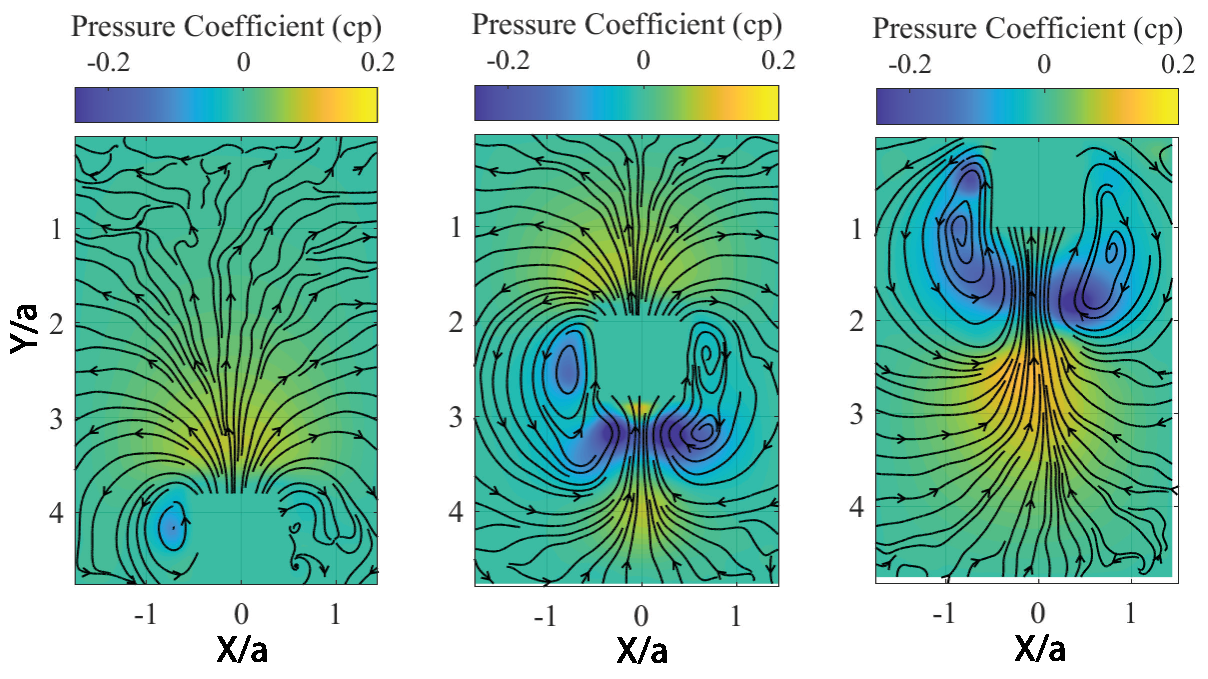}
        \caption{}
        \label{fig:flow_V250}
    \end{subfigure}\\
    \begin{subfigure}[b]{0.72\textwidth}
        \includegraphics[width=\textwidth]{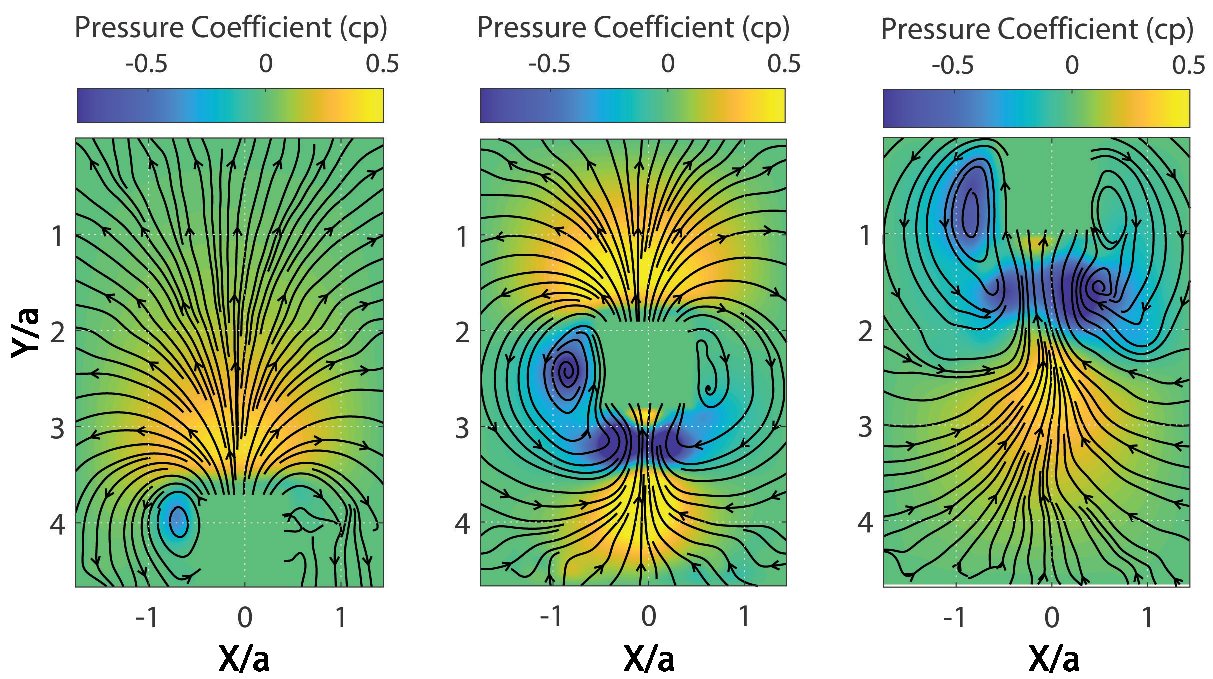}
        \caption{}
        \label{fig:flow_V600}
    \end{subfigure}\\
    \begin{subfigure}[b]{0.72\textwidth}
        \includegraphics[width=\textwidth]{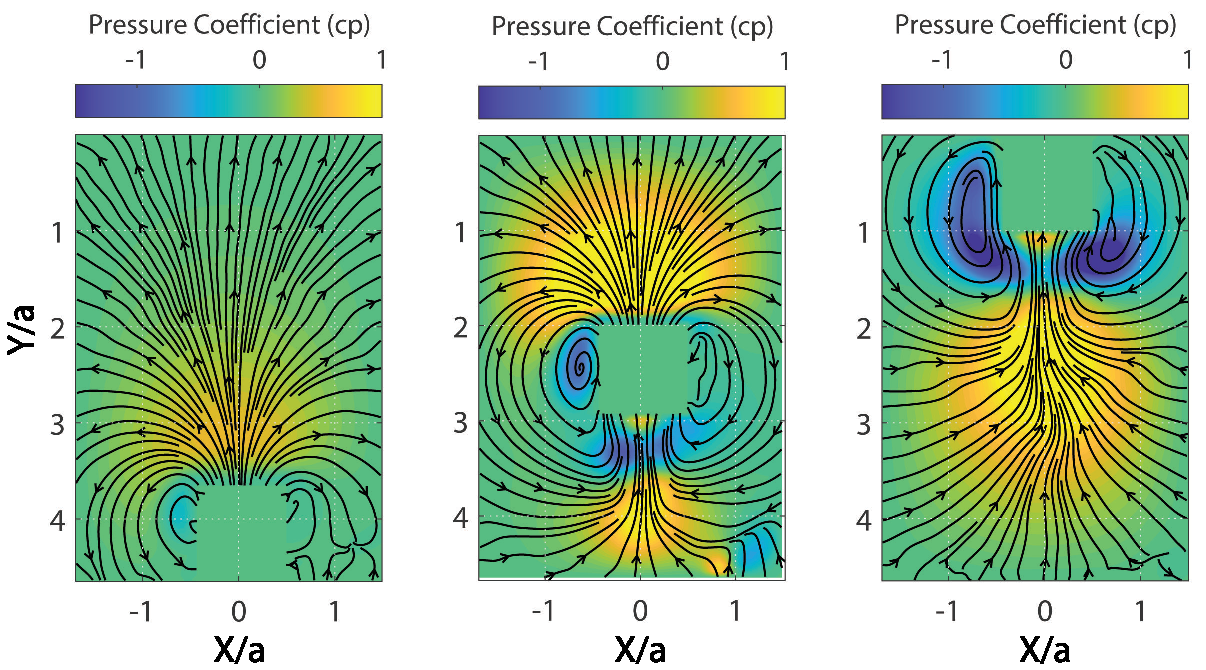}
        \caption{}
        \label{fig:flow_V1000}
    \end{subfigure}
    \caption{Flow structure during exit at (a) $\mathrm{Fr}=0.16$, (b) $\mathrm{Fr}=0.90$, (c) $\mathrm{Fr}=2.54$.}
    \label{fig:flowstructure}
\end{figure*}

Figure~\ref{fig:flowstructure} shows instantaneous velocity and pressure fields at three representative Froude numbers. Across the tested range, the wake is dominated by a single counter–clockwise (CCW) vortex attached to the cylinder base, without evidence of periodic shedding. The corresponding low–pressure region expands with increasing $\mathrm{Fr}$, while a stagnation–induced high–pressure zone develops near the cylinder’s top surface. 

At low $\mathrm{Fr}<0.3$, the recirculation bubble is weak and confined, resulting in minimal surface deformation and entrainment.  
For moderate $0.3 \leq \mathrm{Fr} \leq 1.0$, the CCW vortex intensifies and the wake enlarges, coinciding with stronger free–surface deformation.  
At high $\mathrm{Fr}>1.0$, the negative–pressure zone grows substantially, but the wake remains non-periodic, with no vortex shedding observed.

\subsection{Shear--vortex balance from Okubo--Weiss, swirl strength evolution and $\Lambda$}
\label{subsec:ShearVortexBalance}

\begin{figure*}
    \centering
    \begin{subfigure}[b]{0.72\textwidth}
        \includegraphics[width=\textwidth]{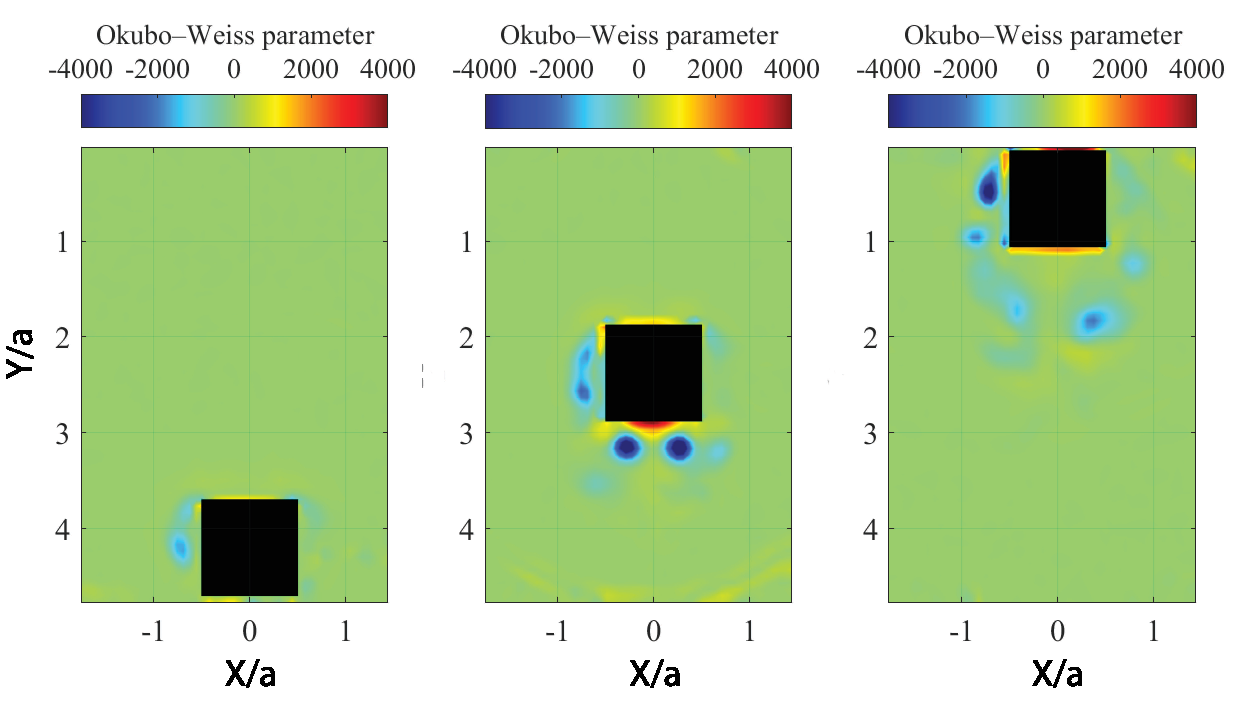}
        \caption{}
        \label{fig:ow_V250}
    \end{subfigure}\\
    \begin{subfigure}[b]{0.72\textwidth}
        \includegraphics[width=\textwidth]{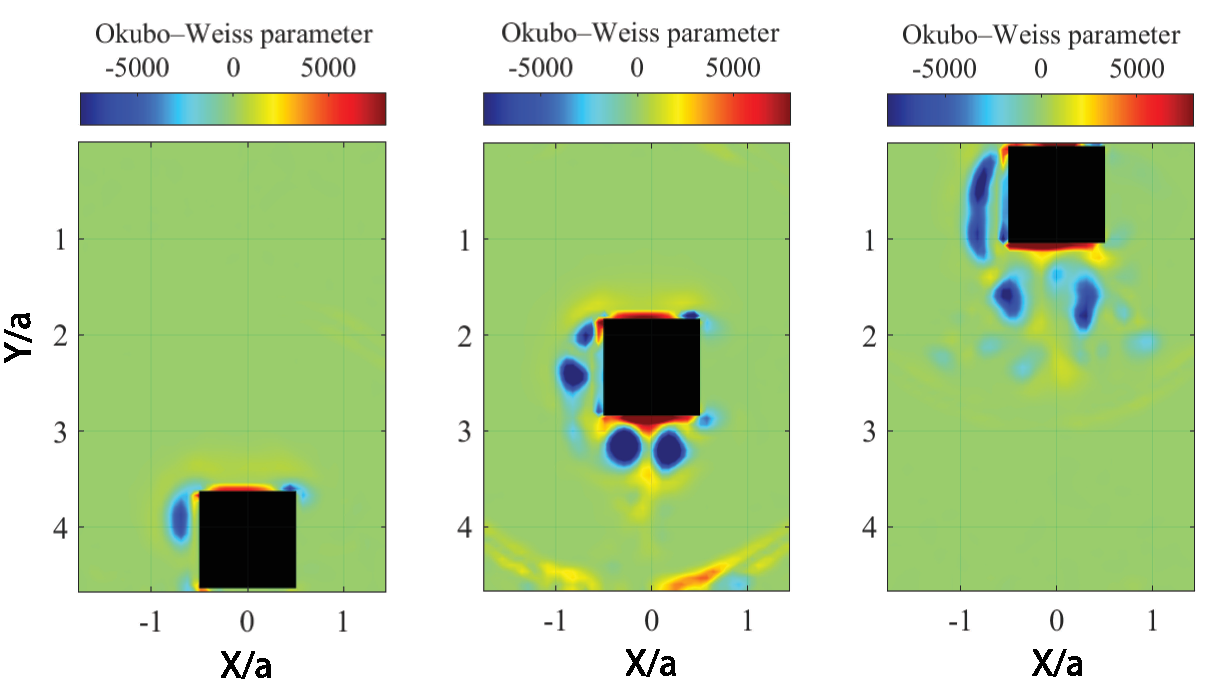}
        \caption{}
        \label{fig:ow_V600}
    \end{subfigure}\\
    \begin{subfigure}[b]{0.72\textwidth}
        \includegraphics[width=\textwidth]{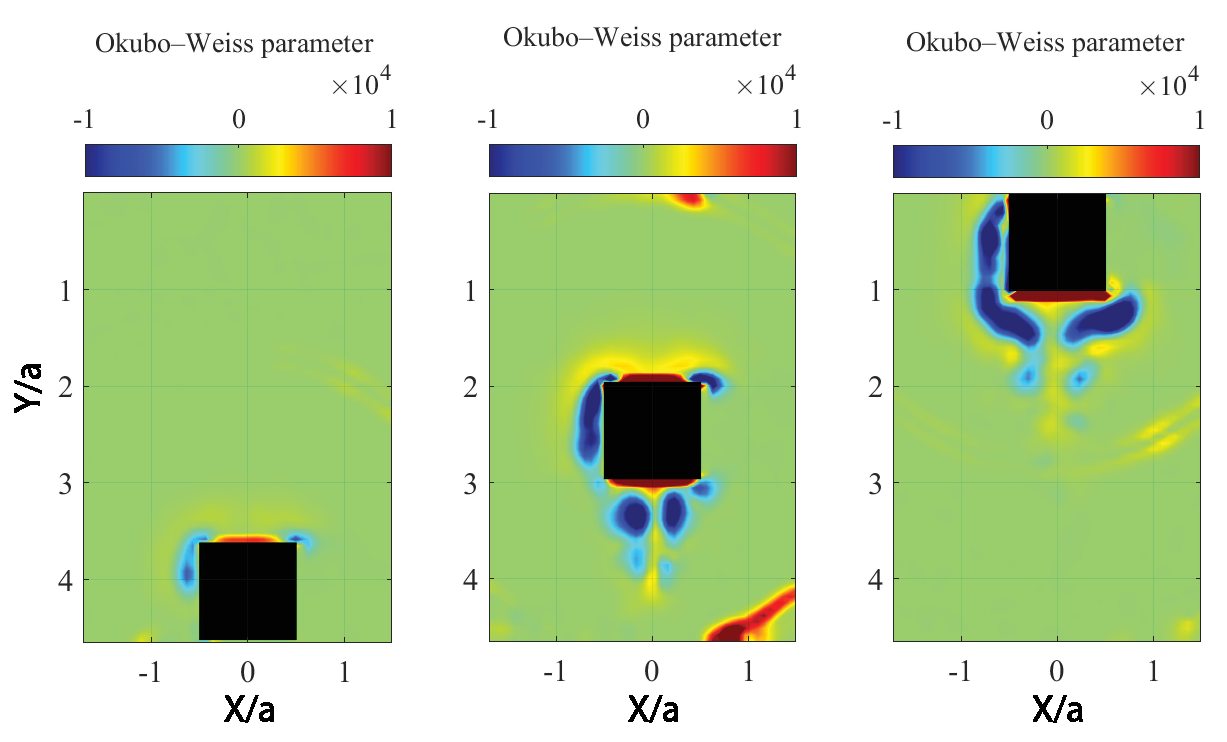}
        \caption{}
        \label{fig:ow_V1000}
    \end{subfigure}
    \caption{Okubo--Weiss parameter $W$ during exit at (a) $\mathrm{Fr}=0.16$, (b) $\mathrm{Fr}=0.90$, (c) $\mathrm{Fr}=2.54$. Negative values ($W<0$) denote rotation–dominated zones.}
    \label{fig:ow_maps}
\end{figure*}

The Okubo--Weiss maps (Fig.~\ref{fig:ow_maps}) show that the wake remains rotation--dominated ($W<0$) under all conditions, consistent with the persistence of a single attached vortex pair. With increasing $\mathrm{Fr}$, the area of negative-$W$ regions expands, confirming the growing influence of rotational motion.

\textcolor{black}{The mean swirl strength $\lambda_{ci}$ (Fig.~\ref{fig:Lamda1}\subref{fig:lambda_m}) increases monotonically as the cylinder rises toward the free surface (decreasing $h$), reflecting progressive intensification of the counter-rotating vortex pair during ascent. At large submergence ($h > 3$), the vortex cores form gradually from weak shear layers and remain spatially diffuse, resulting in low local rotational intensity. As the cylinder approaches the surface, the shear layers roll up progressively into tighter, more concentrated vortices, concentrating rotational motion into smaller spatial domains and raising $\lambda_{ci}$.}

\textcolor{black}{Notably, the curves exhibit a two-slope behavior with a transition near $h \approx 0.9$. At deeper submergence ($h > 1$), $\lambda_{ci}$ increases gradually, corresponding to the slow vortex roll-up phase. Closer to the surface ($h < 0.9$), the slope steepens significantly, indicating rapid vortex intensification as free-surface interaction compresses and organizes the vortex pair. This transition coincides with the onset of observable surface deformation and mirrors the two-regime circulation behavior reported in Fig.~\ref{fig:Circ1}. The Froude number exerts a dominant effect on swirl strength magnitude.}

The shear--vortex interaction parameter $\Lambda$ provides a complementary view (Fig.~\ref{fig:Lamda1}\subref{fig:Lambda_h}). At low $\mathrm{Fr}$, $\Lambda$ stays nearly constant and below unity, indicating that vortex cores occupy more area than shear layers. At moderate to high $\mathrm{Fr}$, $\Lambda$ develops a local maximum around $h \approx 2$. \textcolor{black}{The local maximum in the shear–vortex interaction parameter $\Lambda$ at intermediate submergence ($h \approx 2$) arises from the transient balance between shear-layer and vortex-core regions during the cylinder's ascent. At large $h$ (deep submergence), the flow is still adjusting as shear layers separate from the cylinder's sharp corners, and vortex cores have not yet fully formed. Because rotation-dominated regions ($W < 0$) occupy a small footprint at this stage, the denominator in $\Lambda$ remains small, keeping the ratio low. As the cylinder rises, the shear layers begin to roll up into coherent counter-rotating vortices. During this transition phase near $h \approx 2$, both shear-dominated ($W > 0$) and rotation-dominated ($W < 0$) regions occupy comparable spatial extents, maximizing their ratio and producing the peak in $\Lambda$. This corresponds to the most intense interaction between the nascent vortex cores and the surrounding shear layers. Closer to the free surface ($h < 1$), the vortex pair strengthens and rotation-dominated regions expand substantially, as confirmed by the swirl-strength and pressure-field data. The shear layers become subordinate to the dominant vortex cores, causing $\Lambda$ to decrease and approach a rotation-dominated regime. This progression—from shear-layer formation through transient interaction to final vortex dominance—reflects the natural evolution of the wake and has been documented in classical cylinder-wake studies. The Froude-number dependence of the peak location reflects the accelerated or delayed vortex formation timescales at different exit speeds.} As the cylinder approaches $h=0.5$, $\Lambda$ decreases again, showing that rotational cores dominate the flow topology in the near–surface region.

\textcolor{black}{At low Froude numbers ($Fr \lesssim 0.3$), the shear--vortex interaction parameter $\Lambda$ remains above unity and increases monotonically during the early ascent (Fig.~\ref{fig:Lambda_h}). This indicates that near-wall shear layers dominate 
over rotational motion, consistent with a shear-dominated regime where the weak inertial forcing suppresses coherent vortex roll-up. 
This interpretation aligns with the Okubo--Weiss analysis discussed in Section~\ref{subsec:OWparameter}, where positive $W$ values similarly denote regions dominated by strain or shear, while negative values correspond to vorticity-dominated zones. As the Froude number increases beyond $Fr \approx 0.3$, enhanced inertial forcing promotes the roll-up of separated shear layers, leading to a transient maximum in $\Lambda$ around $h \approx 2$ and a progressive transition toward a vortex-dominated wake. This transition mirrors the shift from viscous–shear control to inertially driven vortex formation reported previously for water-exit flows~\cite{ashraf2024}}.

\subsection{Vortex intensity: swirl strength and circulation}
\label{subsec:SwirlCirc}

Circulation $\Gamma$ provides an integrated measure of vortex strength (Fig.~\ref{fig:Circ1}\subref{fig:Gamma_vs_h}). For all $\mathrm{Fr}$, $\Gamma$ grows as the cylinder approaches the free surface ($h \approx 0.5$) and \textcolor{black}{then tends to decrease. This decline arises from the combined effects of free-surface interaction and vortex dissipation during the final stage of exit. As the cylinder nears the surface, the low-pressure vortex cores begin to deform and spread due to surface bulging and the breakdown of the pressure gradient that sustains coherent rotation. The free surface acts as a no-stress boundary, allowing vorticity to diffuse laterally and vertically, reducing the integrated circulation within the measurement plane. Additionally, as the cylinder body approaches and breaches the interface, the vortex pair detaches and begins to dissipate, no longer fed by continuous shear-layer roll-up from the moving boundary. This drop is therefore a signature of vortex weakening and detachment, not a saturation of vortex strength.} 

\begin{figure*}
    \centering
    \begin{subfigure}[b]{0.49\textwidth}
        \includegraphics[width=\linewidth]{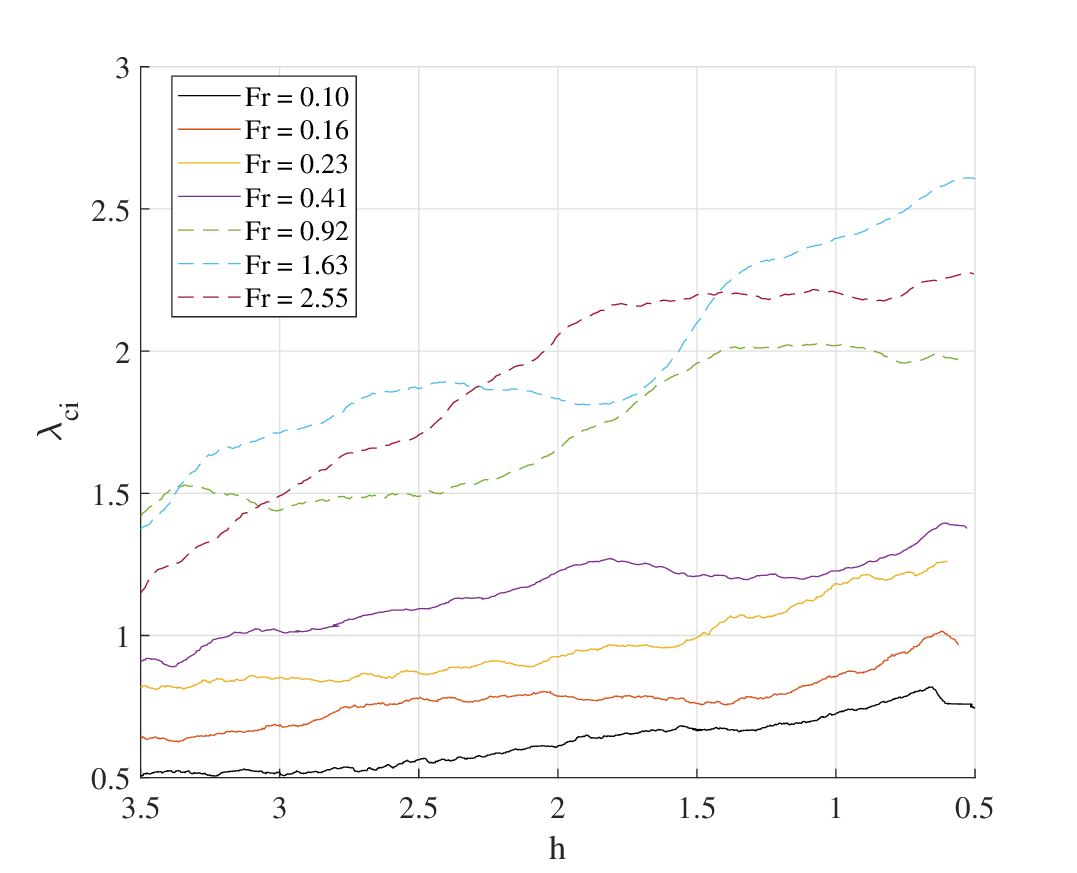}
        \caption{}
        \label{fig:lambda_m}
    \end{subfigure}\hfill
    \begin{subfigure}[b]{0.49\textwidth}
        \includegraphics[width=\linewidth]{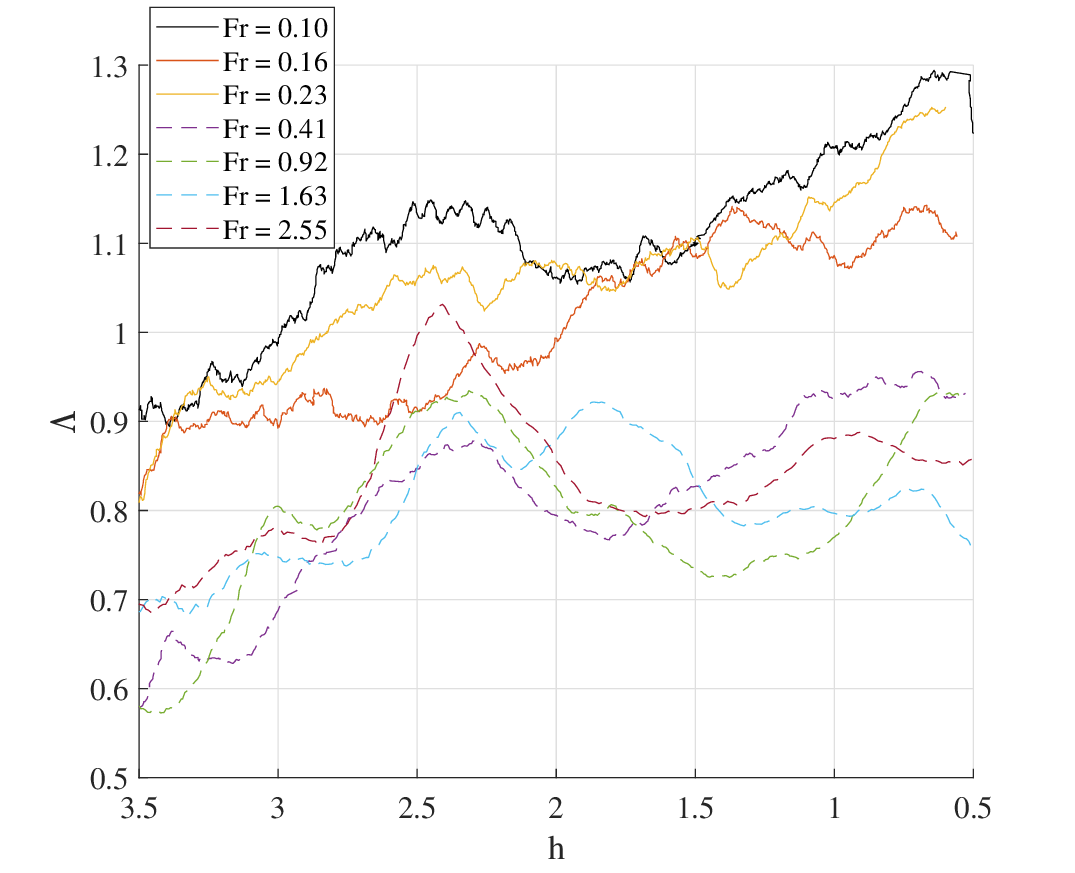}
        \caption{}
        \label{fig:Lambda_h}
    \end{subfigure}
    \caption{(a) Mean swirl strength $\lambda_{ci}$ versus submergence $h$ and (b) Shear--vortex interaction parameter $\Lambda$ versus $h$ for several $\mathrm{Fr}$.}
    \label{fig:Lamda1}
\end{figure*}

\begin{figure*}
    \centering
    \begin{subfigure}[b]{0.49\textwidth}
        \includegraphics[width=\linewidth]{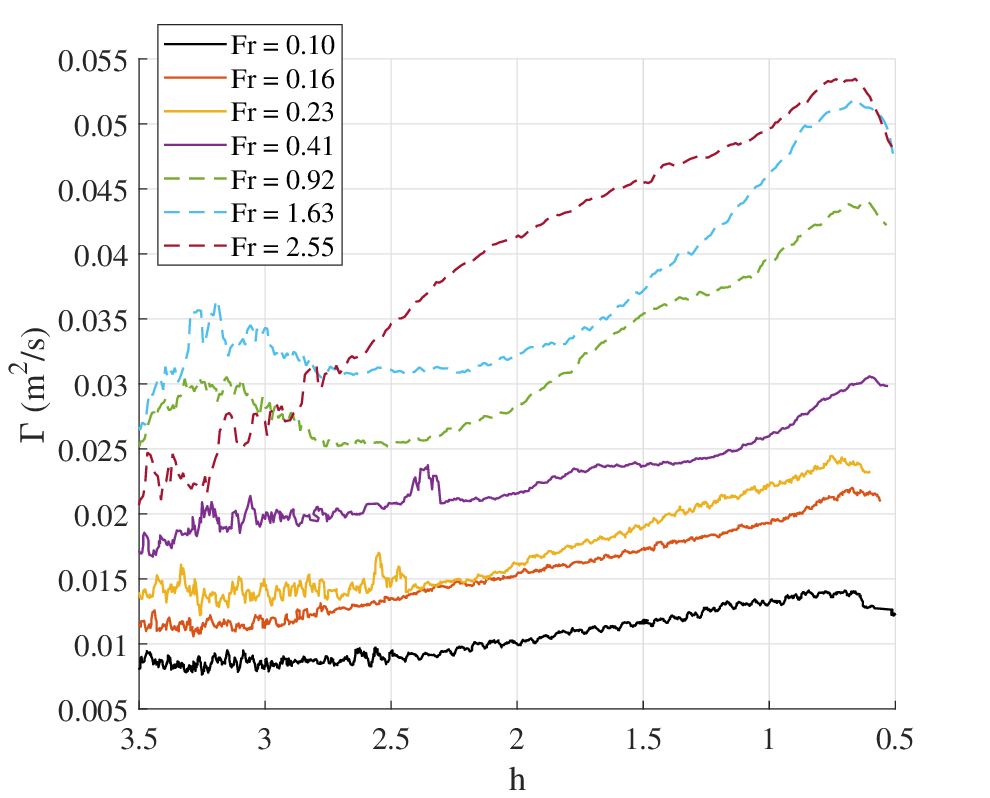}
        \caption{}
        \label{fig:Gamma_vs_h}
    \end{subfigure}\hfill
    \begin{subfigure}[b]{0.49\textwidth}
        \includegraphics[width=\linewidth]{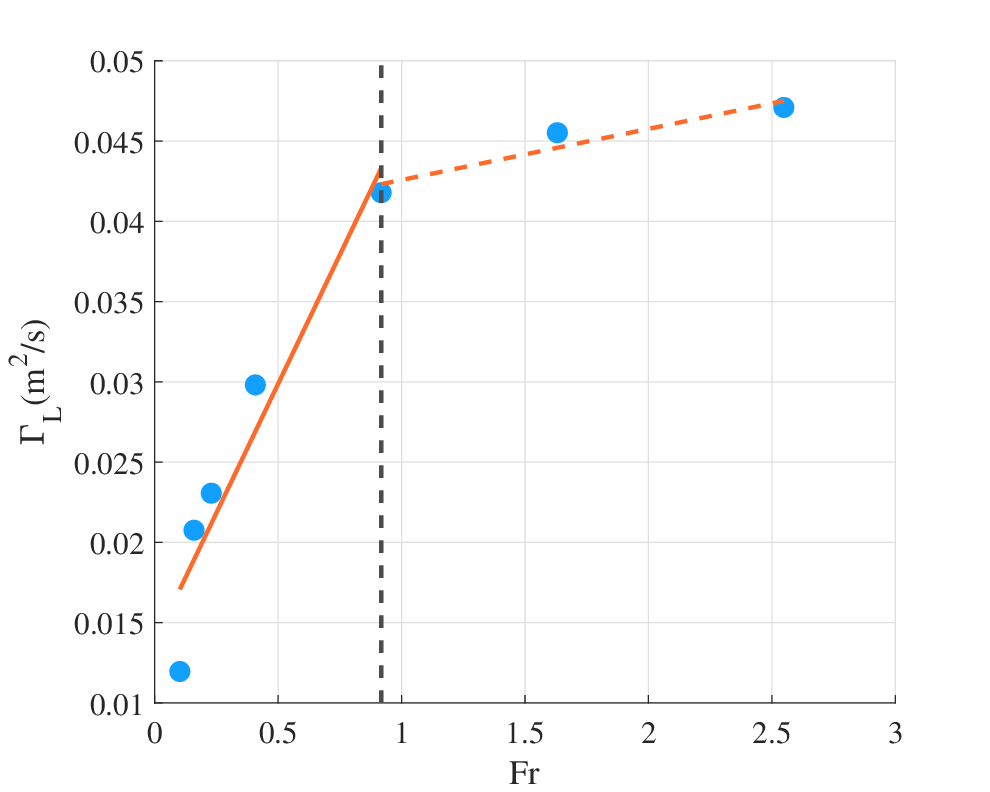}
        \caption{}
        \label{fig:Gamma_at_h05_vs_Fr}
    \end{subfigure}
    \caption{(a) Circulation $\Gamma$ of the CCW vortex versus submergence $h$ and (b) Circulation $\Gamma_L$ at $h=0.5$ versus $\mathrm{Fr}$.}
    \label{fig:Circ1}
\end{figure*}

\textcolor{black}{Fig.~\ref{fig:Circ1}\subref{fig:Lambda_h}, reveals the two-regime circulation response by plotting the circulation $\Gamma_L$ measured at $h = 0.5$ (near the surface) as a function of Froude number. This relationship directly links vortex strength to exit velocity and clarifies the mechanism underlying the entrainment force scaling reported previously\cite{ashraf2024exit}.} 

\textcolor{black}{ At low Froude numbers (Fr $< 0.9$), circulation increases steeply and approximately linearly with Fr. In this regime, the cylinder exits slowly enough that the vortex pair forms gradually, and vortex strength scales directly with the velocity gradients generated at the cylinder's sharp corners. The shear layers separate cleanly, roll up efficiently, and produce coherent vortices whose intensity is proportional to the imposed motion. Each increment in exit speed translates almost directly into increased circulation, reflecting a viscosity-dominated, quasi-steady formation process.}

\textcolor{black}{ At higher Froude numbers (Fr $> 0.9$), the slope decreases significantly, and circulation approaches a saturation plateau. In this regime, faster exit speeds no longer produce proportionate increases in vortex strength. This saturation arises because at high speeds, free-surface deformation becomes pronounced, and the low-pressure vortex cores begin to interact strongly with the deforming interface. The free surface acts as a stress-free boundary that dissipates vorticity and limits further circulation growth. Additionally, rapid motion reduces the time available for vortex organization, so even though velocity gradients are stronger, the integrated circulation cannot continue to grow at the same rate. This transition near Fr $\approx 0.9$ matches the critical Froude number identified in the entrainment force measurements, confirming that the two-regime force behavior is governed directly by vortex circulation dynamics. The saturation indicates that beyond a certain speed, hydrodynamic loading is controlled by vortex intensity limits imposed by free-surface interaction, not by further velocity increases.}

\subsection{Vortex size and entrainment link}
\label{subsec:AreaEntrainment}

\begin{figure*}
    \centering
    \includegraphics[scale=0.55]{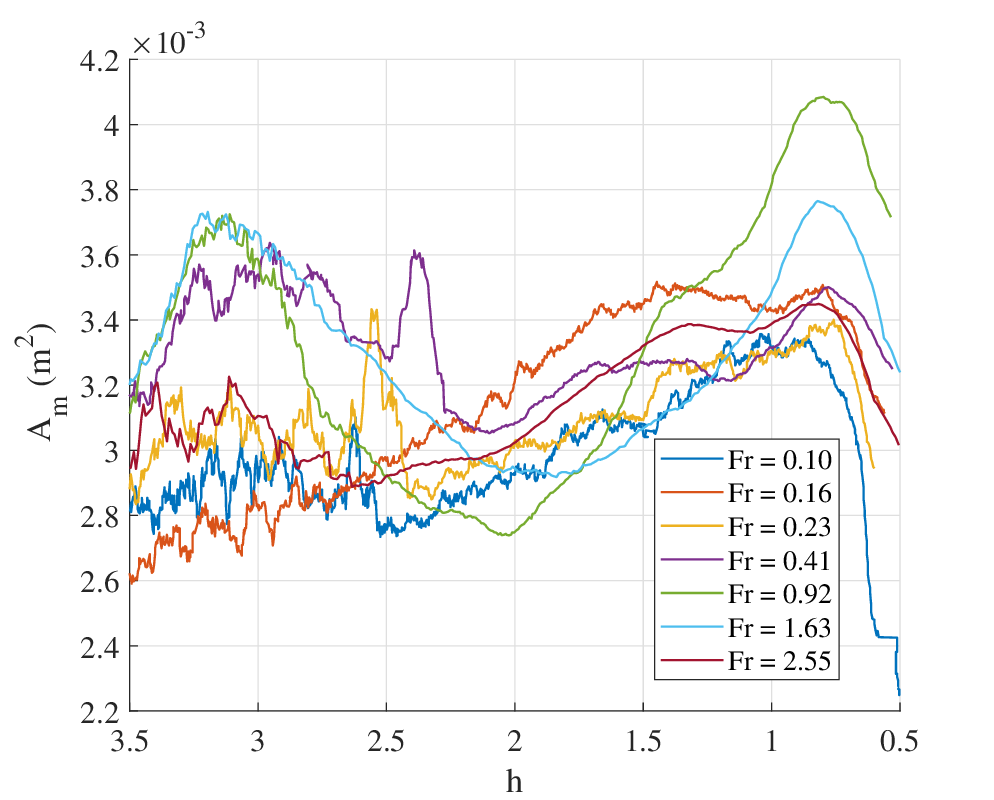}
    \caption{Area $A$ of the CCW vortex versus submergence $h$ at different $\mathrm{Fr}$.}
    \label{fig:Area}
\end{figure*}

\textcolor{black}{The Fig.~\ref{fig:Area} shows vortex area plot as a function of Froude number, $Fr$, and cylinder's position, $h$.  There is an observable peak in vortex area $A$  near intermediate submergence ($h \approx 1.5$--$0.5$). It might arise from the onset of free-surface interaction. At large $h$, the vortex pair develops freely and expands as shear layers roll up. As the cylinder approaches the surface, the low-pressure region in the vortex cores begins to deform the free surface, creating the surface bulge observed in visualizations. Before strong interaction sets in, the vortex reaches its maximum spatial extent.  Closer to the surface ($h < 1$), the free-surface interaction intensifies, compressing and distorting the vortex pair and reducing its effective area. The Froude-number dependence reflects the timing of this interaction: at higher Fr, the cylinder moves faster, and surface interaction begins earlier (at larger $h$), shifting the peak accordingly. This interpretation is consistent with the pressure-field analysis (Fig.~\ref{fig:flowstructure}) and the observed surface deformation at high Fr.
}

However, the vortex area $A$ (Fig.~\ref{fig:Area}) shows only weak dependence on $\mathrm{Fr}$ and remains nearly constant compared with circulation. This indicates that the enhanced entrainment observed at higher $\mathrm{Fr}$ is driven primarily by increased rotational intensity rather than by a larger vortex footprint. The maximum circulation $\Gamma$ at $h=0.5$ scales with the entrainment force, highlighting a direct dynamical link between vortex strength and the fluid volume carried above the free surface.

\section{Conclusions}
\label{sec:conclusions}

We have experimentally investigated the exit dynamics of a square cylinder rising through quiescent water across a range of Froude numbers $\mathrm{Fr}$, combining time–resolved PIV with complementary vortex diagnostics (swirl strength $\lambda_{ci}$, Okubo–Weiss parameter $W$, and the shear–vortex interaction metric $\Lambda$). This multi–diagnostic framework establishes a consistent physical picture of how shear, rotation, and entrainment evolve during water exit.

A central result is the persistence of a counter–rotating vortex pair that remains attached to the cylinder throughout its ascent. Unlike entry or cross–flow wakes, no von Kármán–type shedding emerges, even at high $\mathrm{Fr}$. The dominance of $W<0$ regions confirms that the wake topology is rotation–dominated rather than shear–dominated, explaining the absence of periodic vortex release. This finding identifies exit wakes as a distinct canonical problem in unsteady wake dynamics.

We also uncover a two–regime circulation response: for $\mathrm{Fr}<1$, vortex strength and entrainment increase steeply, while for $\mathrm{Fr}>1$, the growth rate saturates. This transition aligns with previously reported entrainment–force scaling \cite{ashraf2024exit}, establishing circulation as the direct dynamical link between coherent vortices and the fluid volume carried across the free surface. Importantly, the vortex area remains nearly constant with $\mathrm{Fr}$, indicating that entrainment is governed primarily by rotational intensity, not footprint size.

Finally, the $\Lambda$ diagnostic reveals a depth–dependent balance: shear–layer influence peaks at intermediate submergence ($h \approx 2$), while near the free surface ($h \to 0.5$) the wake transitions toward pure vortex dominance. This clarifies how shear–vortex interaction modulates entrainment during the approach to the interface.

In summary, the upward translation of a square cylinder produces a robust, non–shedding wake where intensified rotation, rather than vortex enlargement, drives entrainment. The identification of a circulation–based two–regime transition around $\mathrm{Fr}\approx 1$ advances the fundamental understanding of exit wakes and provides a framework for linking hydrodynamic loads to coherent vortex dynamics. Beyond practical implications for naval and offshore design, these results contribute to the broader physics of vortex–interface interactions in unsteady multiphase flows.

\section*{Acknowledgements}

The financial support of the Belgian Fund for Scientific Research under research project WOLFLOW (F.R.S.-FNRS, PDR T.0021.18) is gratefully acknowledged. Part of the experimental setup was financed by {\it Fonds Sp\'eciaux} from ULi\`ege. SD is F.R.S--FNRS Senior Research Associate

\section*{Authors’ Contributions}

IA conceived and designed the study, performed the experiments, processed the PIV data, developed the vortex identification framework, carried out the analysis, and wrote the manuscript.  

SD supervised the research, provided guidance on experimental design and interpretation, and contributed to refining the discussion and conclusions.  

NT assisted with the development of the pressure–estimation code and provided technical input during the analysis and methodology development.

\section*{Appendices: Additional Flow-Field Visualizations and Okubo–Weiss Maps}

\begin{figure*}
	\centering
	\includegraphics[scale=00.7]{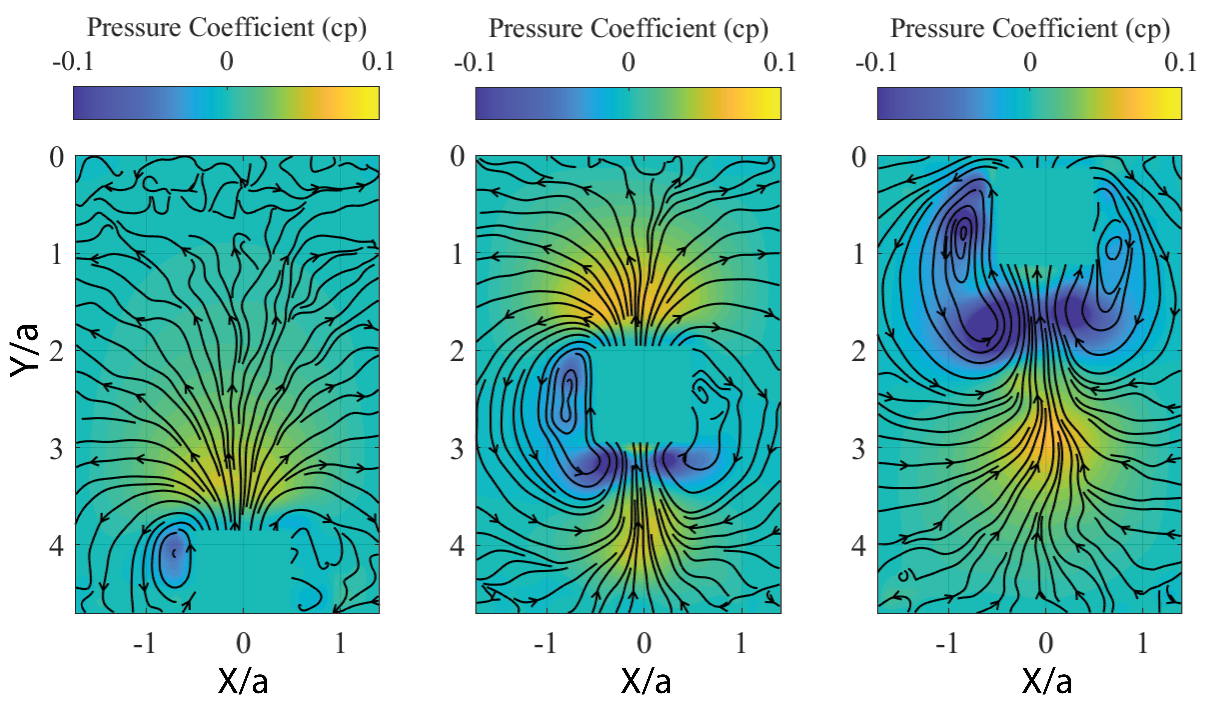}
	\caption{Flow structure during the exit of a square cylinder at Froude number, Fr= 0.10.}
	\label{V200}
\end{figure*}

\begin{figure*}
	\centering
	\includegraphics[scale=00.7]{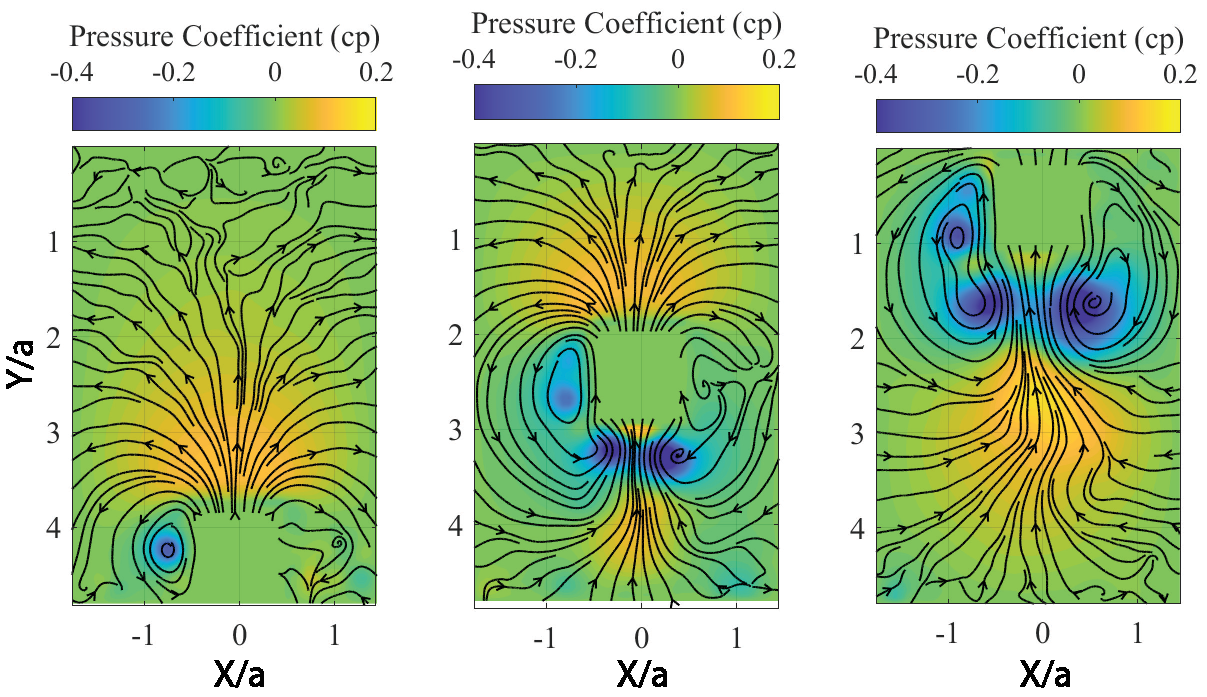}
	\caption{Flow structure during the exit of a square cylinder at Froude number, Fr= 0.23}
	\label{V300}
\end{figure*}

\begin{figure*}
	\centering
	\includegraphics[scale=00.7]{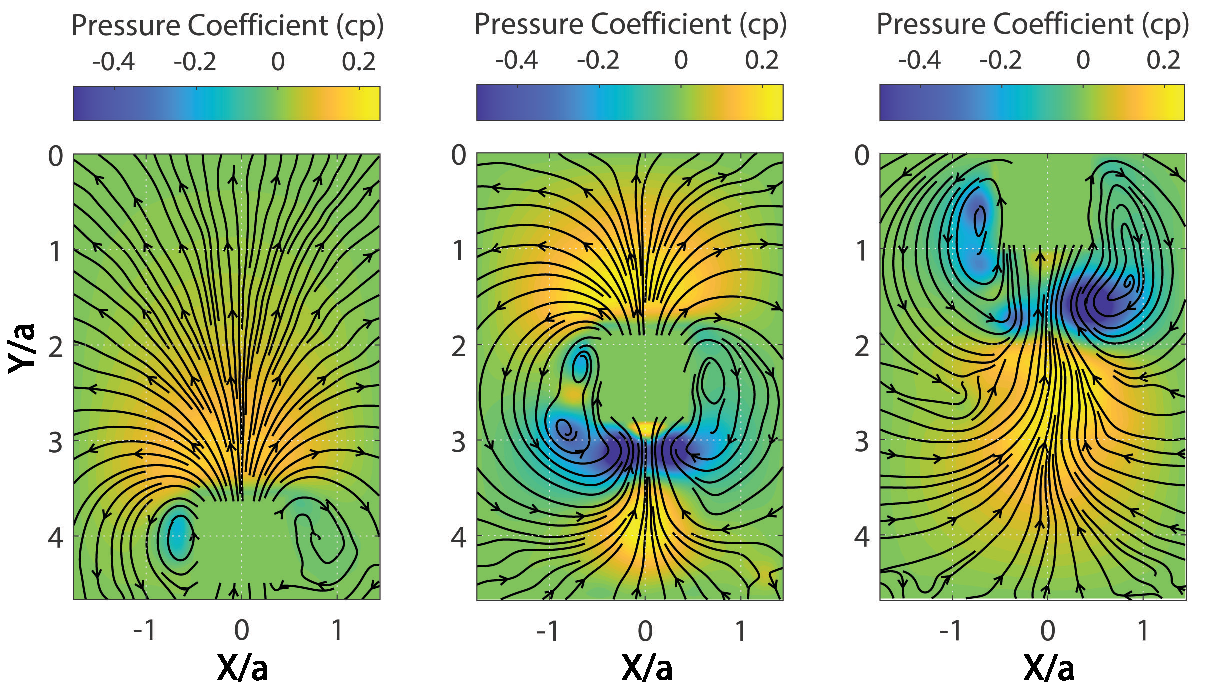}
	\caption{Flow structure during the exit of a square cylinder at Froude number, Fr= 0.40}
	\label{V400}
\end{figure*}

\begin{figure*}
	\centering
	\includegraphics[scale=00.7]{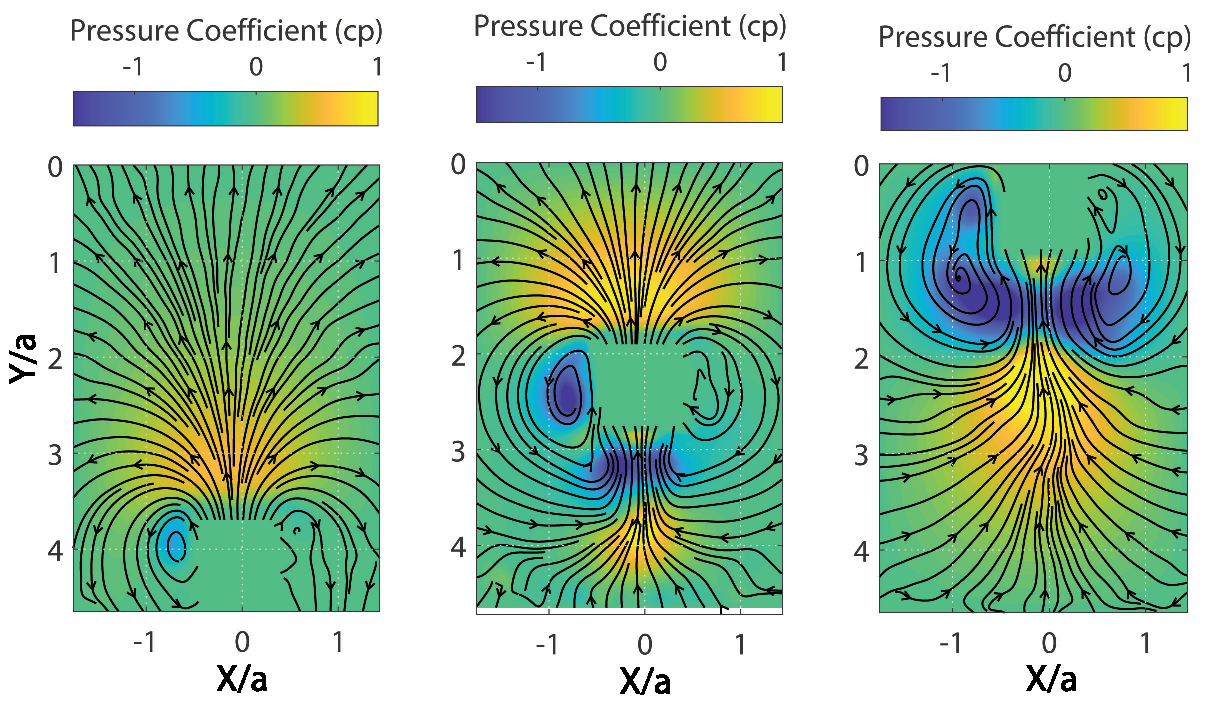}
	\caption{Flow structure during the exit of a square cylinder at Froude number, Fr = 1.61}
	\label{V800}
\end{figure*}

\begin{figure*}
	\centering
	\includegraphics[scale=00.7]{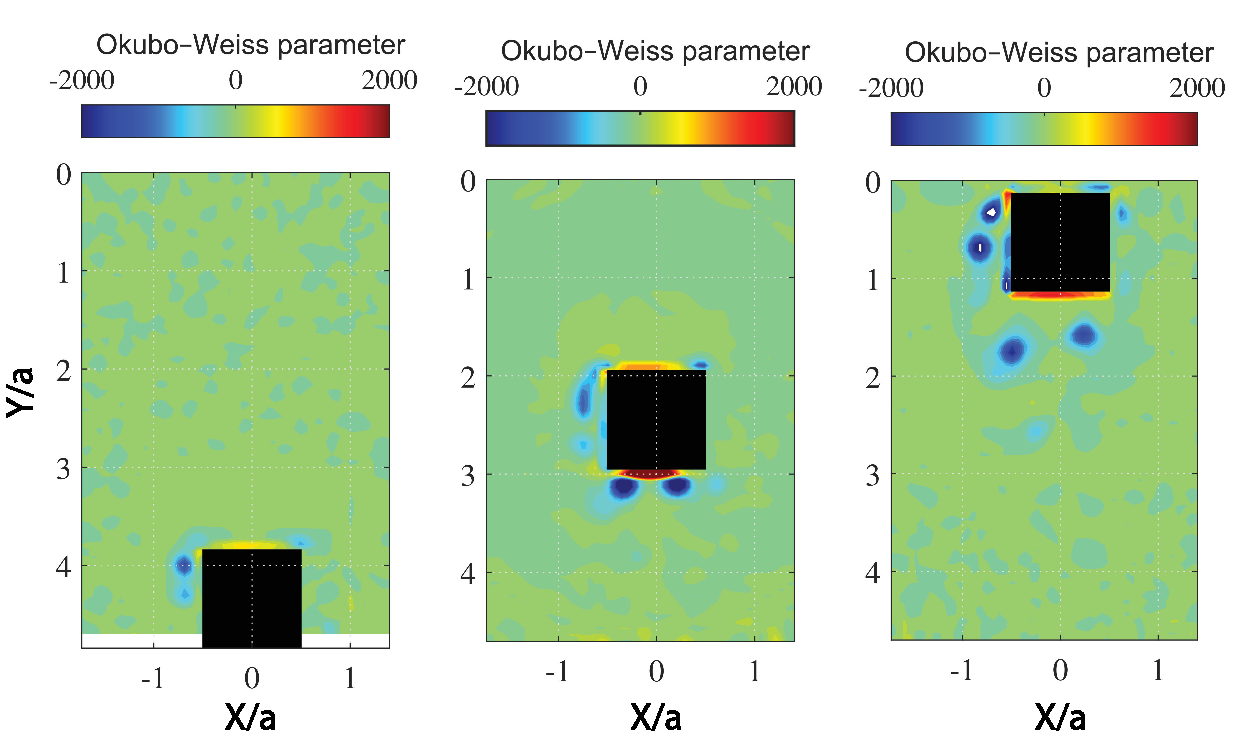}
	\caption{Okubo–Weiss parameter (W) during the exit of a square cylinder at Froude number, Fr= 0.10.}
	\label{V200_0}
\end{figure*}

\begin{figure*}
	\centering
	\includegraphics[scale=00.7]{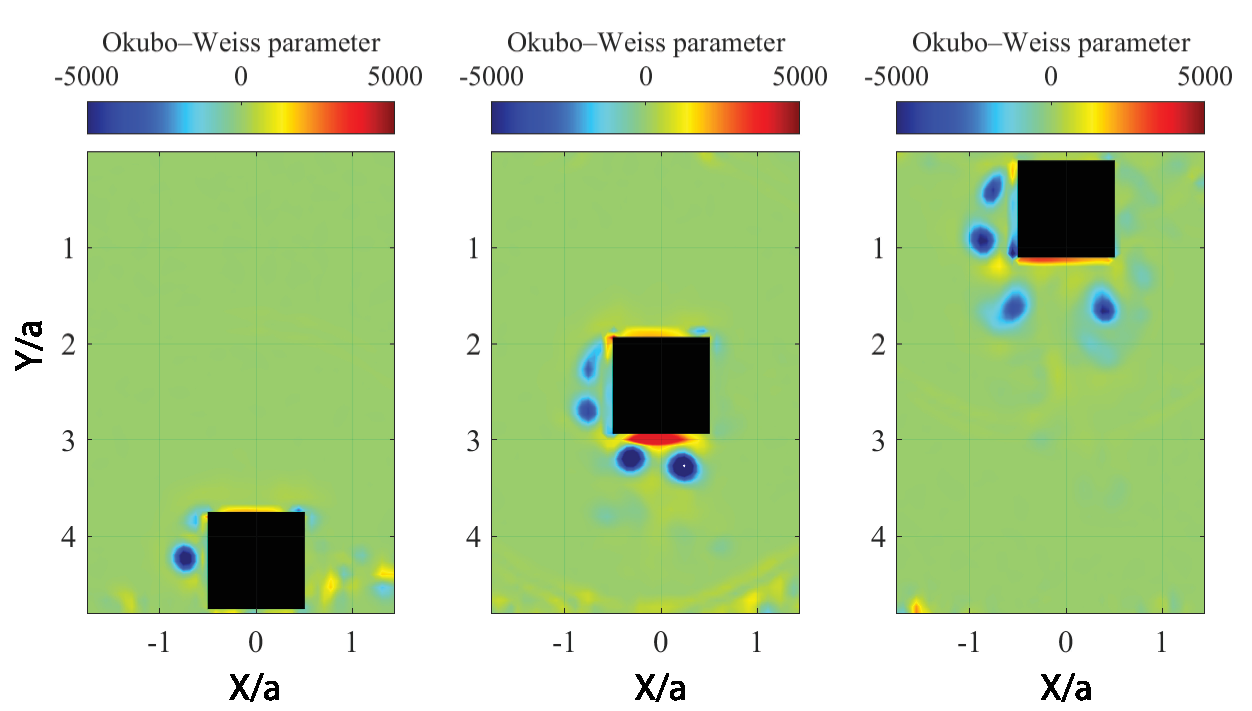}
	\caption{Okubo–Weiss parameter (W) during the exit of a square cylinder at Froude number, Fr= 0.23}
	\label{V300_0}
\end{figure*}

\begin{figure*}
	\centering
	\includegraphics[scale=00.7]{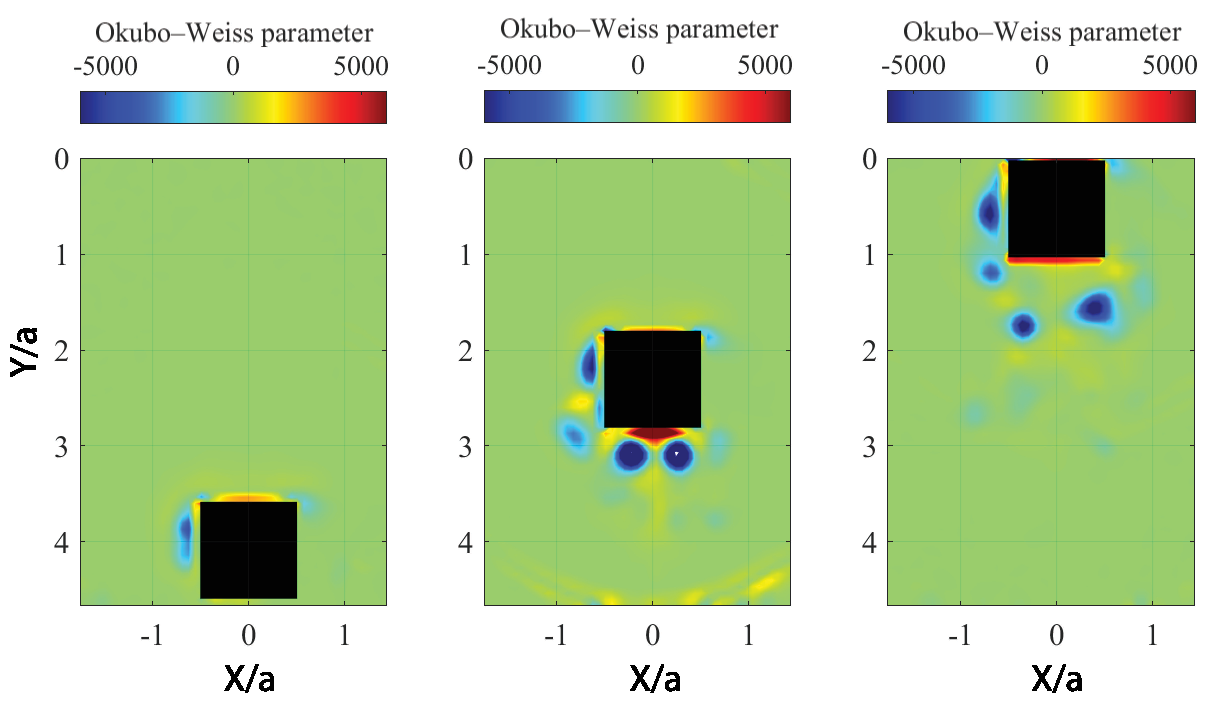}
	\caption{Okubo–Weiss parameter (W) during the exit of a square cylinder at Froude number, Fr= 0.40}
	\label{V400_0}
\end{figure*}

\begin{figure*}
	\centering
	\includegraphics[scale=00.7]{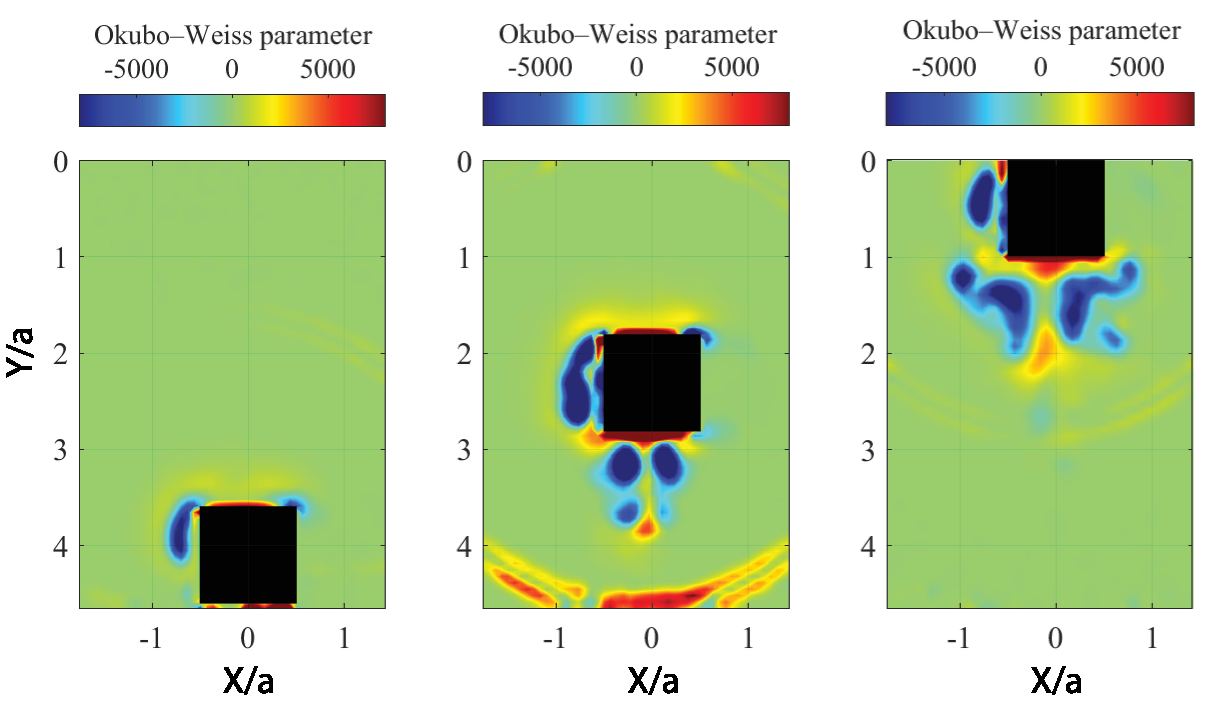}
	\caption{Okubo–Weiss parameter (W) during the exit of a square cylinder at Froude number, Fr = 1.61}
	\label{V800_0}
\end{figure*}

\clearpage

\bibliography{aipsamp}

\end{document}